\documentclass[twocolumn,prb,showpacs,multicol,amsmath,amssymb]{revtex4}
\usepackage[dvips]{graphicx}
\usepackage{graphicx}
\usepackage{dcolumn}
\usepackage{bm}
\usepackage{graphics}
\usepackage{epsfig,color,soul}

\newcommand{\be}{\begin{equation}}
\newcommand{\ee}{\end{equation}}
\newcommand{\bea}{\begin{eqnarray}}
\newcommand{\eea}{\end{eqnarray}}

\begin{document}
\title{Signature of the quantum critical point on the witness of the non-Markovianity and the entanglement in spin-1/2 chain}
\author{Z. Saghafi}
\author{S. Samadi Rezaei}
\author{P. Azizi}
\author{E. Hosseini Lapasar}
\author{S. Mahdavifar}
\affiliation{Department of Physics, University of Guilan, 41335-1914, Rasht, Iran}
\date{\today}
\begin{abstract}

A thermodynamic limit chain of spin-1/2 particles with $XX$ and three-spin interactions (TSI) is considered. Using the fermionization technique, the Hamiltonian of the chain with periodic boundary conditions is exactly diagonalized. In the ground-state phase diagram of the chain system, a quantum critical point separates the Luttinger liquid and chiral phases. Selecting one-spin as an open quantum system, the rest of the spins play the role of its environment. By choosing different initial states, we studied the dynamics of the entanglement between the open quantum system and its environment. 
In the initial state, the state of the open quantum system is a superposition of up and down states and the environment is polarized. The revival of the entanglement is observed in the whole range of interactions as an indicator of non-Markovian dynamics. Our exact results revealed that non-Markovian behavior is independent of the initial state. By tuning the initial state, the open quantum system will be completely entangled with the environment at a special time called entanglement-time, $t_E$. The value of the mentioned entanglement-time decreased by increasing TSI. The signature of a quantum critical point is confirmed by investigating the long-time average of the entanglement and using the trace-distance as a witness of non-Markovianity.

\end{abstract}
\maketitle

\section{Introduction}\label{sec1}

Open quantum system dynamics has raised a great deal of interest both theoretically and experimentally \cite{Eberly07, Yu09, Werlang09, Mazzola10, Amico04, Horodecki09}.
Using the theory of open quantum systems, the dynamical behavior of the system which interacts with its environment is classified into two categories: Markovian and non-Markovian \cite{Breuer16}. The Markovian process involves the flow of information from the system to the environment. On the other hand, the process by which part of the information is returned from the environment to the system is called non-Markovian dynamics. Describing the memory effects in the non-Markovian process is a fundamental subject in the theory of open systems \cite{Breuer2}. Investigating dynamics of the open quantum system was initially allocated to the Markovian behavior, it has recently expanded to non-Markovian behaviors \cite{Piilo08, Madsen11, Liu11, Huelga12, Barnes12, Addis14, Orieux15}. Systems that are in contact with different non-Markovian environments show the revival of entanglement\cite{ Bellomo07, Franco12, Franco13}. It is essential to show the process of the revival of entanglement to introduce a general measurement of a non-Markovianity degree in open quantum systems \cite {Breuer09}. Note that in Markovian dynamics, the concurrence decreases exponentially \cite{Santos06}. \\
There are several ways to identify non-Markovianity via witnesses. A witness of non-Markovianity is a quantity which will be vanished for all Markovian dynamics \cite{Chru12, Chru14}. We know that a non-zero value of a witness of non-Markovianity shows a non-Markovian dynamics. The witnesses of non-Markovianity can be classified into two categories: the witnesses based on monotonicity under completely positive maps and the witnesses based on monotonicity under local completely positive maps \cite{Rivas14}.
The first category includes the following methods: Trace distance and the BLP quantifier \cite {Breuer09, Laine10}, Fidelity \cite {Uhlmann76, Jozsa94}, Quantum relative entropies \cite {Lindblad75}, Quantum Fisher information \cite {Holevo82, Braunstein94}, Capacity measure \cite {Bylicka13}, Bloch volume measure \cite {Lorenzo13} and the second one contains Entanglement \cite {Plenio7}, Quantum mutual information \cite {Luo12}, and Quantum discord \cite {Zurek00, Ollivier1, Modi12}. 

Theoretically, in a recent work, the dynamics of a qubit coupled to a spin chain environment has been studied \cite {Apollaro11}. The environment is described by a XY model in the transverse magnetic field. The dynamics of the qubit is Markovian at a special point. Two regions that are separated by this point are non-Markovian and lead to two completely different dynamical behaviors. It has been shown that the contribution of energy density is responsible for the non-Markovian effects. This result in an infinite environment is also correct. From an experimental point of view, a non-Markovianity-assisted high-fidelity refined Deutsh-Jozsa algorithm is implemented with a solid spin in a diamond\cite{Dong18}. Especially, a non-Markovian quantum process is observed by measuring the non-Markovianity of the spin system. The control of the degree of non-Markovianity in the dynamics of a nitrogen-vacancy center electron spin is also demonstrated experimentally \cite{Hass18}. They have shown that,  by changing the population of the nitrogen spin, the non-Markovianity of the electron spin’s dynamics is tuneable.  In addition, using a randomized set of central radio-frequency fields, a non-Markovian environment for a single nuclear magnetic resonance qubit is effectively realized \cite{Khurana19}. The study of the role of different interactions in inducing non-Morkovianity has been going on in recent years\cite{Lorenzo11, Znidaric11, Haikka12, Dubertrand18, Saghafi19-1, Saghafi19-2, Motamedifar19, Chakraborty19, Roy19, Matern19, Roos19, Kyaw20}.

Recently, systems involving multiple interactions, such as three-spin interaction, four-spin interaction, etc., are significant by various parts of physics \cite { Titvinidze03, Lou04, Lou06, Taras08, Tahvili17, Shadman18, Moshfegh19, Yin20, Hu20}. A wide range of spin-1/2 Hamiltonians can be created in different configurations of an optical lattice \cite {Kuklov03}. One type of multiple-spin interaction is a three-spin interaction (TSI) which can be represented by a triangular configuration \cite { Duan03}. TSI has been proposed with various Hamiltonians, including the following
\begin{eqnarray}
{\cal{H}}_{TSI}^{(1)}&=&J'\sum_{j=1}^{N} (\vec{S}_j~.~\vec{S}_{j+2})_{xy}  ~ {S}_{j+1}^z, \nonumber\\
{\cal{H}}_{TSI}^{(2)}&=&J'\sum_{j=1}^{N}[ (\vec{S}_j~ \times  ~\vec{S}_{j+2})~.~ \hat{k} ]~ {S}_{j+1}^z.
\label{TSI}
\end{eqnarray}

Thermodynamic properties and quantum phase transition for both interactions with the Heisenberg spin-1/2 XX model have been investigated \cite { Titvinidze03, Lou04}. There is a significant difference between the results of these two models. 
Some of the differences express: The model that containing ${\cal{H}}_{TSI}^{(1)}$, shows that it has spontaneous magnetization in its ground state, while the second model has zero magnetization in its ground state. In the study of the specific heat as a function of $T/J$, it is observed that the first model has a two-peak structure, while the second model has a single-peak structure. 

The dynamics of entanglement and dynamical transition from the Markovian to the non-Markovian regime for a one-dimensional spin-1/2 XX model with ${\cal{H}}_{TSI}^{(1)}$ has been studied \cite { Mahmoudi17}. In this work, we focus on the dynamical behavior for the system involving ${\cal{H}}_{TSI}^{(2)}$ to explore the interplay between the critical phenomena, quantum correlations, and non-Markovianity. One spin is selected from the chain and it plays the role of the open system, while the rest of the chain constitutes its environment. A class of initial states is considered, where the environment is polarized and the one-spin system is a superposition of up and down states.  Using the fermionization approach, Hamiltonian is diagonalized and exact expression for the time-dependent entanglement between the one-spin system and its environment is obtained. Revivals phenomenon of entanglement is showed that the dynamics is non-Markovian in presence of the TSI interaction.  A similar result is obtained by calculating the trace distance.  For initial states where one-spin is also polarized, the open quantum system will be completely entangled with the environment at a special time called entanglement-time, $t_E$. Entanglement time decreases by increasing TSI. On the other hand, we also calculate the long-time average of the entanglement and will find a clear signature of the quantum critical point. 

The paper is structured as follows. In section II, first, the model is introduced, then using the fermionization technique, the Hamiltonian is diagonalized and analytical results are given. In section III, the dynamics of entanglement between one-spin and its environment is studied. Results on the trace distance are reported in section IV. Finally, we conclude all of our results in section V.

\section{THE MODEL}\label{sec2}

The Hamiltonian of a spin-1/2 XX model with ${\cal{H}}_{TSI}^{(2)}$ is given by
\begin{eqnarray}
{\cal H}&=&-J \sum_{j=1}^{N}({S}_{j}^x {S}_{j+1}^x+{S}_{j}^y {S}_{j+1}^y)  \nonumber\\
&-&J'\sum_{j=1}^{N} ({S}_{j}^x {S}_{j+2}^y-{S}_{j}^y {S}_{j+2}^x )~ {S}_{j+1}^z, 
\label{Hamiltonian s1}
\end{eqnarray}
where $S_{j}$ is the spin-1/2 operator on the $j$-th site, $J$ and $J'$ are respectively the exchange coupling between the spins on the nearest neighbors sites and on the three nearest neighbors sites. $J'=0$ represents the isotropic XX model. 
The TSI breaks $\pi/2$-rotation symmetry along the $z$-axis. Since the dynamics of the system is governed by Hamiltonian, so we expect remarkable changes in the dynamics that will occur.

The Hamiltonian can be exactly diagonalized by the fermionization approach. In the first step, using the Jordan-Wigner transformation 
\begin{eqnarray}
{S}_{j}^{+}&=&a_{j}^{\dagger} \exp(i\pi\sum_{l<j}a_{l}^{\dagger}a_{l}),\\
{S}_{j}^{-}&=&a_{j} \exp(-i\pi\sum_{l<j}a_{l}^{\dagger}a_{l}),\\ \
{S}_{j}^{z}&=&a_{j}^{\dagger}a_{j}-\frac{1}{2}, 
\label{J-W transformation}
\end{eqnarray}
the  Hamiltonian is mapped onto a 1D model of noninteracting spinless fermions
\begin{eqnarray}
{\cal H}_{f}&=& \frac{-J}{2} \sum_{j=1}^{N} (a^{\dag}_{j}a_{j+1}+a^{\dag}_{j+1}a_{j})\nonumber \\
&+& i\frac{ J'}{2} \sum_{j=1}^{N} (a^{\dag}_{j}a_{j+2} - a^{\dag}_{j+2}a_{j}).
\end{eqnarray}
In the next step, performing a Fourier transformation into the momentum space as $a^{\dag}_{j} = \frac{1}{\sqrt{N}} \sum ^{N} _{j=1} e^{ikj} a^{\dag}_{k}$, the diagonalized Hamiltonian is given  by
\begin{eqnarray}
{\cal H}_{f}=\sum_{k}\varepsilon(k) a_{k}^{\dagger} a_{k},
\label{Hamiltonian d1}
\end{eqnarray}
where $\varepsilon(k)$ is the dispersion relation
\begin{eqnarray}
\varepsilon(k) &=& -J(\cos(k)-\frac{\alpha}{2} \sin(2 k)),
\label{spec}
\end{eqnarray}
with $\alpha=\frac{J'}{J}$. The ground state phase diagram is well known\cite{Lou04}. System is in the gapless Luttinger liquid phase in the region $\alpha \leq \alpha_c=1$ and shows chiral ordering with gapless excitations  in the region $\alpha > \alpha_c=1$.  

Here, we are interested to study the dynamical behavior of a one-spin open quantum system and understanding how the TSI changes the dynamical behavior of the system. To study the dynamics and measuring the degree of non-Markovianity, we use the trace distance, hence we need to calculate the reduced density matrix. In the first step, it is necessary to determine the initial state of the system at $t=0$. Then by using the time evolution operator ($\hbar=1$ is considered) as 
\begin{eqnarray}
U(t)=e^{-it\sum_{k}\varepsilon(k) a_{k}^{\dagger} a_{k}/\hbar},
\label{t-evolution}
\end{eqnarray}
the physical state of the system at any time $t$ is obtained. The trace distance which gives a measure of the distinguishability between two initial quantum states ($\rho_1$ and $\rho_2$)  can be expressed as\cite { Breuer09} 
\begin{eqnarray}
D(\rho_1,\rho_2)=\frac{1}{2} \mathrm{tr} \vert \rho_1 - \rho_2 \vert,
\label{TD}
\end{eqnarray}
where $\vert \rho_1 - \rho_2 \vert=\sqrt{(\rho_1 - \rho_2)^{\dagger}(\rho_1 - \rho_2)}$. At any time, if the information is returned from the environment to the system, the trace distance will increase; therefore, it is obviously related to existence of the non-Markovianity. 
The witness of the non-Markovianity, ${\cal {N}}$, is defined as
\begin{eqnarray}
{\cal {N}}=max \int _{\sigma>0} \sigma(t,\rho_{1,2}(0)) dt,
\label{non-Markovianity}
\end{eqnarray}
where 
\begin{eqnarray}
\sigma(t,\rho_{1,2}(0))=\frac{d}{dt} D(\rho_1(t),\rho_2(t)),
\label{sigma}
\end{eqnarray}
is the rate of change of the trace distance at time $t$. This computational method is common for calculating the non-Markovianity \cite{Breuer09, Lorenzo13-1, Mahmoudi17, Saghafi19-1}. It is used in different models as: the damped Janes-Cummings model \cite{Breuer09}, the spin-1/2 transverse field XY Heisenberg chain model \cite{Saghafi19-1}, the extended cluster spin-1/2 XX chain model \cite{Mahmoudi17} and the anisotropic spin-1/2 XY chain model \cite{Lorenzo13-1}.   
It is important to note that if the trace distance increases, we can use the rate of change for calculating the witness of non-Markovianity. Otherwise, we cannot use the rate of change of the trace distance for calculating the witness of non-Markovianity.\\

\section{One tangle}\label{sec3}

\begin{figure}[t]
\centerline{\psfig{file=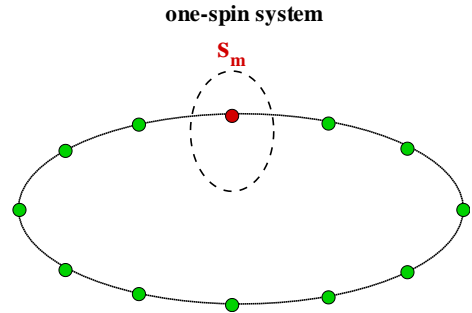,width=2.4 in}}
\caption{ The symbolic form of a one-spin open quantum system and its environment.}
\label{fig1}
\end{figure}

In this section, we select one-spin of the chain as an open quantum system, while the rest of the chain plays the role of the environment. A symbolic form of the system is shown in Fig.~\ref{fig1}. In fact, we concentrate on the entanglement between a single spin located at site $m$ and the rest of the chain system. For this purpose, we consider an initial state where the environment is polarized but the state of the one-spin system is a superposition of up ($|\uparrow\rangle$) and down ($|\downarrow\rangle$) states as 
\begin{eqnarray}
|\psi(t=0)\rangle&=&(\cos(\phi)|\uparrow\rangle+\sin(\phi)|\downarrow\rangle)_S \nonumber \\
& \otimes &|\downarrow\downarrow..... \downarrow\rangle_E,
\label{initial-state of a single-spin}
\end{eqnarray}
which is equivalent to
\begin{eqnarray}
|\psi(t=0)\rangle&=&(\cos(\phi) a_m^{\dag}+\sin(\phi)) ~ |0\rangle.
\label{initial-state of a single-spin}
\end{eqnarray}
in the fermionic form. $\phi$ is a phase factor and $|0\rangle$ denotes the vacuum state. Applying the time evolution unitary operator, the time-dependent physical state of the system is obtained as 
\begin{eqnarray}
|\psi(t)\rangle&=&\frac{1}{\sqrt{N}} ( \sum_{k} e^{i(k m-\varepsilon(k) t)}~ \cos(\phi) a_k^{\dag}\nonumber \\
&+& \sqrt{N} \sin(\phi) ) ~ |0\rangle.
\label{state of a single-spin}
\end{eqnarray}
The one-spin reduced density matrix, $\rho$, is defined as
{\small
\begin{eqnarray}
\rho=\left(
             \begin{array}{cc}
              \frac{1}{2} + \langle{S}_{m}^{z}\rangle & \langle{S}_{m}^{-}\rangle  \\
               \langle {S}_{m}^{+}\rangle & \frac{1}{2} - \langle{S}_{m}^{z}\rangle  \\
               \end{array}
           \right), 
            \label{density matrix1}
\end{eqnarray}
}\\
where $\langle ... \rangle$ denotes the expectation value on the time-dependent physical state. Using the time-dependent state and fermionic operators, we calculated the density matrix as:
{\small
\begin{eqnarray}
\rho=\left(
             \begin{array}{cc}
              A & B  \\
               B^* & 1-A  \\
               \end{array}
           \right), 
            \label{density matrix1}
\end{eqnarray}
}\\
where
\begin{eqnarray}
A(\phi, t)&=&{\cos}^2(\phi)~({f}^2(t)+{g}^2(t)),       \nonumber \\
B(\phi, t)&=& {\sin}(\phi)~ {\cos}(\phi) ~({f}(t)-i {g}(t)),
\end{eqnarray}
and 
\begin{eqnarray}
f(t)&=& \frac{1}{N} \sum_{k} \cos (\varepsilon(k)t), \nonumber \\
g(t)&=& \frac{1}{N} \sum_{k} \sin (\varepsilon(k)t).
\end{eqnarray}
In the following, we concentrate on the entanglement between a single spin and the rest of the chain. In this scenario, the amount of entanglement can be measured by the so-called one-tangle \cite{Amico04}.  It is defined as
\begin{eqnarray}
\tau &=& 4 \det \rho,  \\
&=&4~ {\cos}^4(\phi)~({f}^2(t)+{g}^2(t)) [1-({f}^2(t)+{g}^2(t))]. \nonumber 
\label{tangle}
\end{eqnarray}
At $t=0$, the one-spin system is not entangled with the rest of the chain independent of the phase factor $\phi$. By passing the time, the non-zero value of the entanglement is expected since, at a short time, low-lying excited states contribute to the dynamics of the system. On the other hand, the phase factor $\phi$ does not change the dynamical behavior of the system and only plays the role of a positive constant.
\begin{figure}[t]
\centerline{\psfig{file=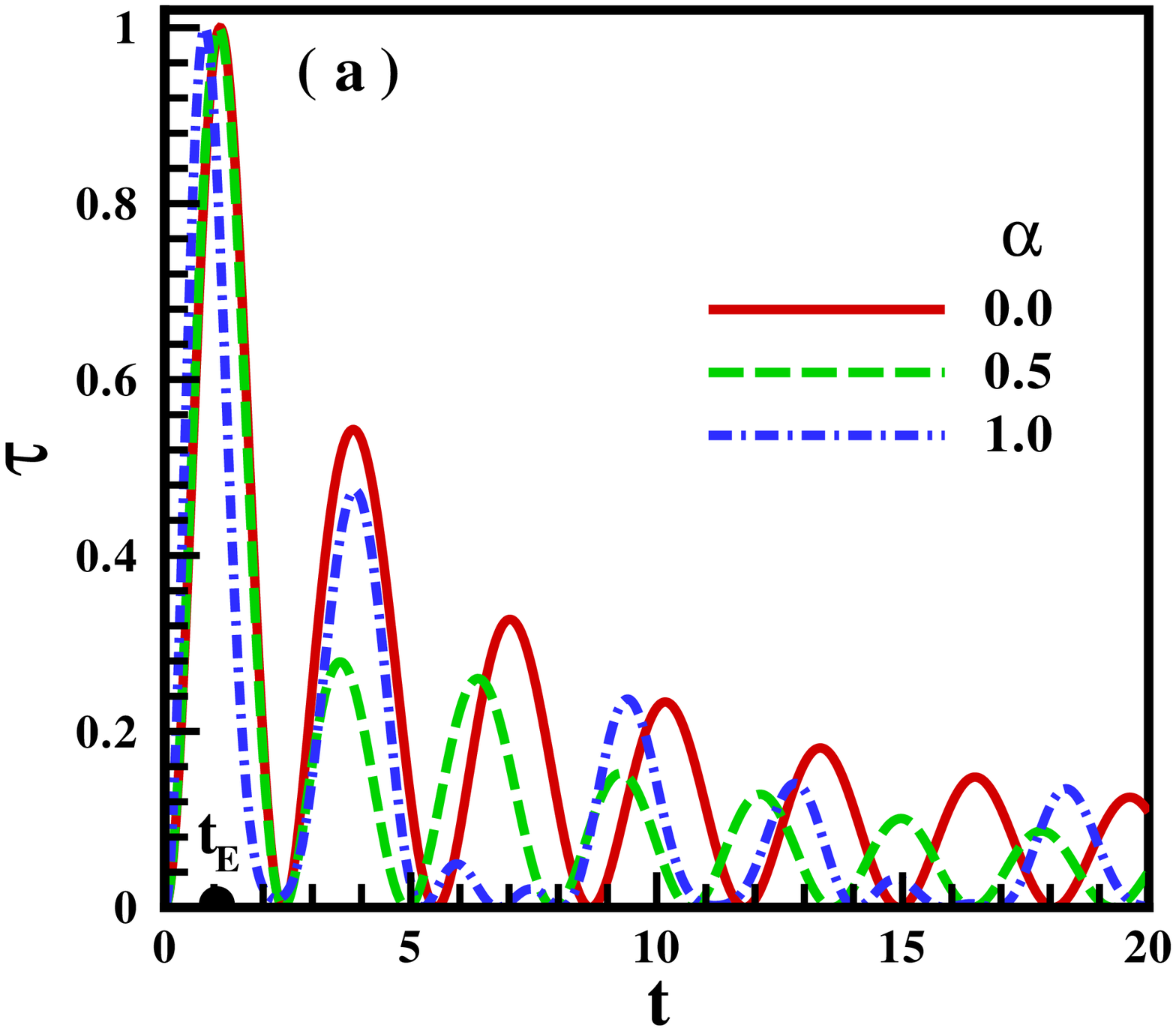,width=1.9 in}\psfig{file=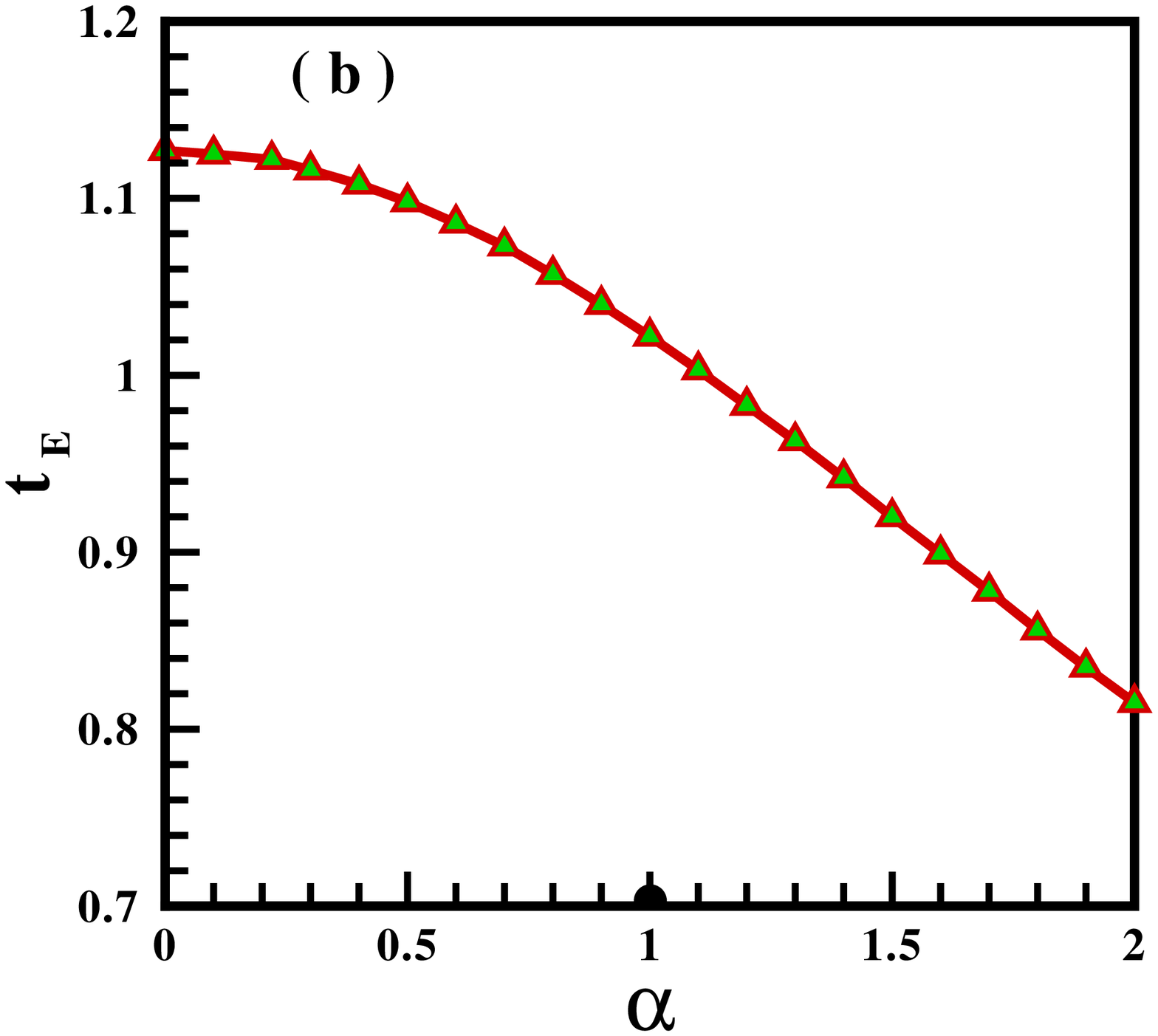,width=1.9 in}}
\centerline{\psfig{file=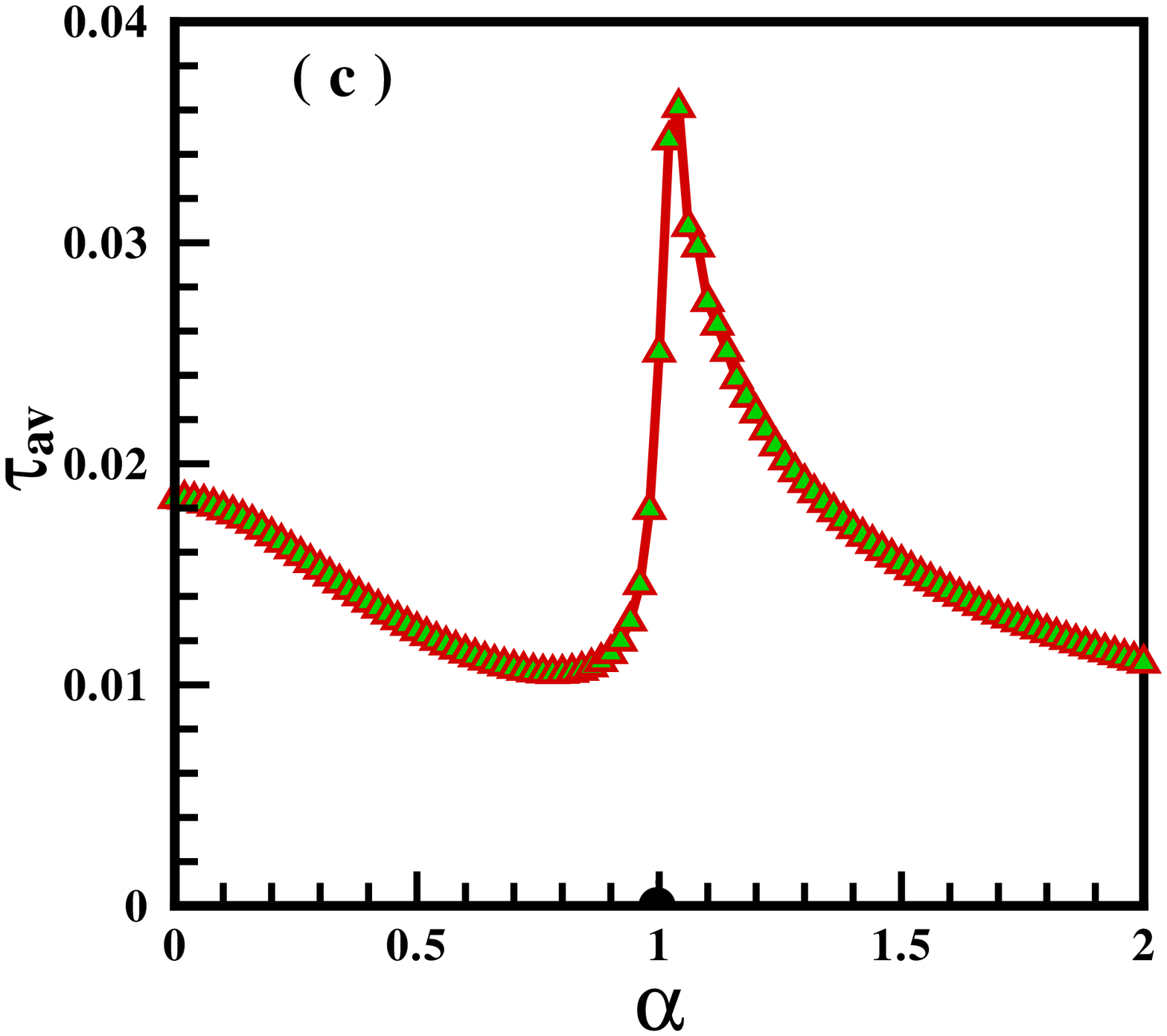,width=1.9 in}}
\caption{(a) The dynamical behavior of the entanglement between the one-spin open quantum system and its environment. (b) The entanglement-time with respect to the TSI. (c) The long-time average of the entanglement with respect to the TSI. }\label{fig22}
\end{figure}

The results presented in Fig.~\ref{fig22} (a) provide information about the dynamical behavior of the entanglement between the one-spin system and the rest of the chain for three different values of the TSI. No entanglement is observed at the initial time $t=0$. There is an upward trend at very short times independent of the value of TSI. The entanglement increases and interestingly the one-spin system will be completely entangled with its environment at a special time $t_E$. It is noticeable that as soon as the time passes from the entanglement-time, $t_E$, the entanglement decreases and reaches zero for the first time.  The mentioned dynamical behavior is repeated periodically by passing time which is known as the revival of the entanglement. In fact, such a revival is due to the reaction of the environment. Feedback of the entanglement from the environment to the open quantum system is one of the special characterizations of the non-Markovianity. We have to mention that the entanglement between system and environment declines moderately and decays at longer times. To find a deeper insight into the nature of the dynamical behavior of entanglement between the one-spin system and its environment, we have also investigated the entanglement-time $t_E$ with respect to the TSI. Results are presented in Fig.~\ref{fig22} (b). The entanglement-time shows monotonic decreasing behavior with respect to the TSI, which can be regarded as an increasing effect of the TSI on the velocity of the entanglement transfer.

\begin{figure}[t]
\centerline{\psfig{file=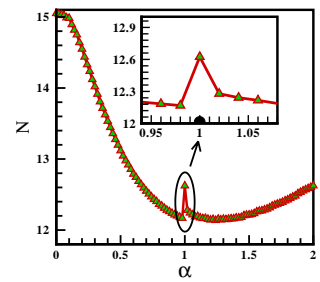,width=1.8 in}}
\caption{The witness of non-Markovianity ${N}$ as a function of TSI. Inset shows results in the quantum critical region.}\label{fig44}
\end{figure}
In addition, we have calculated the long-time average of the entanglement defined as 
\begin{eqnarray}
\tau_{av} = \lim_{T \longrightarrow \infty} \frac{1}{T} \int_0^T \tau(t)~ dt. 
\label{tangle}
\end{eqnarray}
Results are presented in Fig.~\ref{fig22} (c) for different values of TSI. The signature of the quantum critical point is also clearly seen in the behavior of $\tau_{av}$. Three different regions are observed. A quantum critical region, where the TSI is very close to the quantum critical point $\alpha_c=1$ and two regions which are outside of the quantum critical region are distinguished with values of $\alpha<\alpha_c$ and  $\alpha>\alpha_c$. By increasing the TSI from zero, the long-time average of the entanglement decreases slowly up to the edge of the quantum critical region. As soon as the TSI enters into the quantum critical region, $\tau_{av}$ increases rapidly and will be maximized at the second edge of the quantum critical region. In the region, $\alpha>\alpha_c$, it shows monotonic decreasing behavior. 

\section{MEASURE OF NON-MARKOVIANITY}\label{sec3}

In this section, we evaluate the trace distance as a measure of non-Markovianity: we can anticipate similar results as those found for the entanglement. In order to calculate the witness of non-Markovianity, $\rho_1$ and $\rho_2$ must be determined. The two initial states are selected as
\begin{eqnarray}
|\psi_1(t=0)\rangle&=&(\cos(\phi_1)|\uparrow\rangle+\sin(\phi_1)|\downarrow\rangle)_S \otimes(|\downarrow\downarrow..... \downarrow\rangle)_E,\nonumber \\
|\psi_2(t=0)\rangle&=&(\cos(\phi_2)|\uparrow\rangle+\sin(\phi_2)|\downarrow\rangle)_S \otimes(|\downarrow\downarrow..... \downarrow\rangle)_E.\nonumber \\
\label{initial-state of a single-spin}
\end{eqnarray}
Using the time-dependent states and fermionic form, we calculated the $\rho_1-\rho_2$ and finally the trace distance is obtained as 
\begin{eqnarray}
D(\rho_1,\rho_2)&=&\frac{1}{2} \mathrm{tr} \vert \rho_1 - \rho_2 \vert, \nonumber \\
&=&\sqrt{A'^2+|B'|^2}
\end{eqnarray}
where 
\begin{eqnarray}
A'&=&A(\phi_1,t)-A(\phi_2,t), \nonumber \\
B'&=&B(\phi_1,t)-B(\phi_2,t).
\end{eqnarray}
We want to determine whether the dynamical behavior of the one-spin system is Markovian or non-Markovian. To reach this goal, the witness of non-Markovianity ${\cal {N}}$ as a function of the TSI strength is calculated. It should be noted that the time integral is extended over all intervals where trace distance ($D(\rho_1,\rho_2)$) rises by increasing time. Choosing different values of phases $\phi_1$ and $\phi_2$, we are looking to find the maximum distance $\vert \rho_1 - \rho_2 \vert$, which is called the optimal state pair. We have to mention that the optimal state pair leads to the maximal possible backflow from the environment during the time evolution of states. Results of the witness of non-Markovianity ${\cal {N}}$  are shown in Fig.~\ref{fig44}. In fact, the integral in Eq.~(\ref{non-Markovianity}) is calculated numerically for the time interval from $0$ to $1000$ and a time step of $0.01$. It is clearly seen that the dynamics of the system is non-Markovian in the whole range of considered two-point and cluster interactions in complete agreement with the entanglement results. In fact, by increasing TSI from zero, the degree of non-Markovianity ${\cal {N}}$ shows decreasing behavior. As soon as the system placed in the quantum critical region, different behavior is signaled. As is seen in the inset of Fig.~\ref{fig44}, a cusp as a signature of the quantum critical point is observed at $\alpha_c\sim 1$.     
\begin{figure}[t]
\centerline{\psfig{file=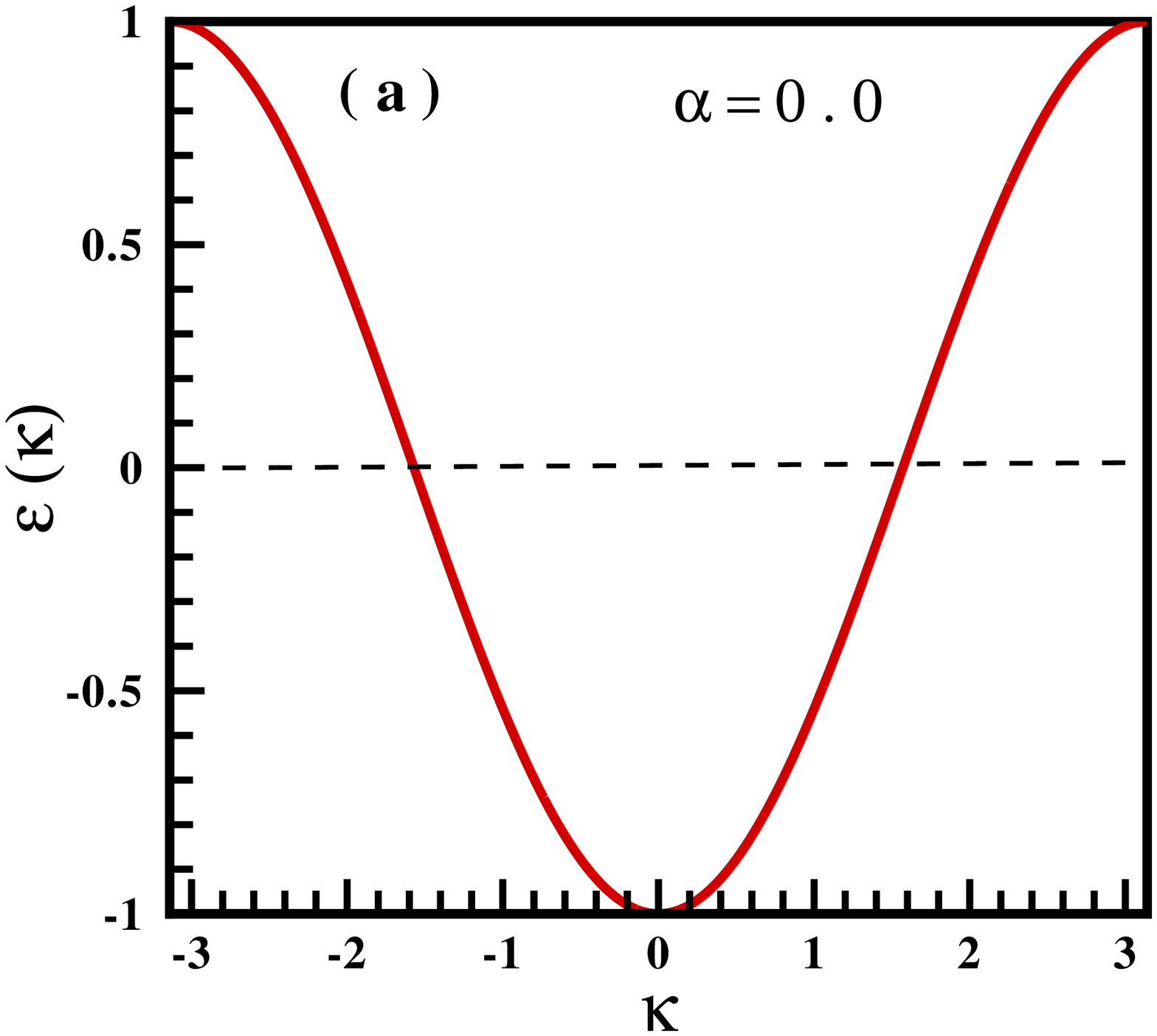,width=1.8 in}\psfig{file=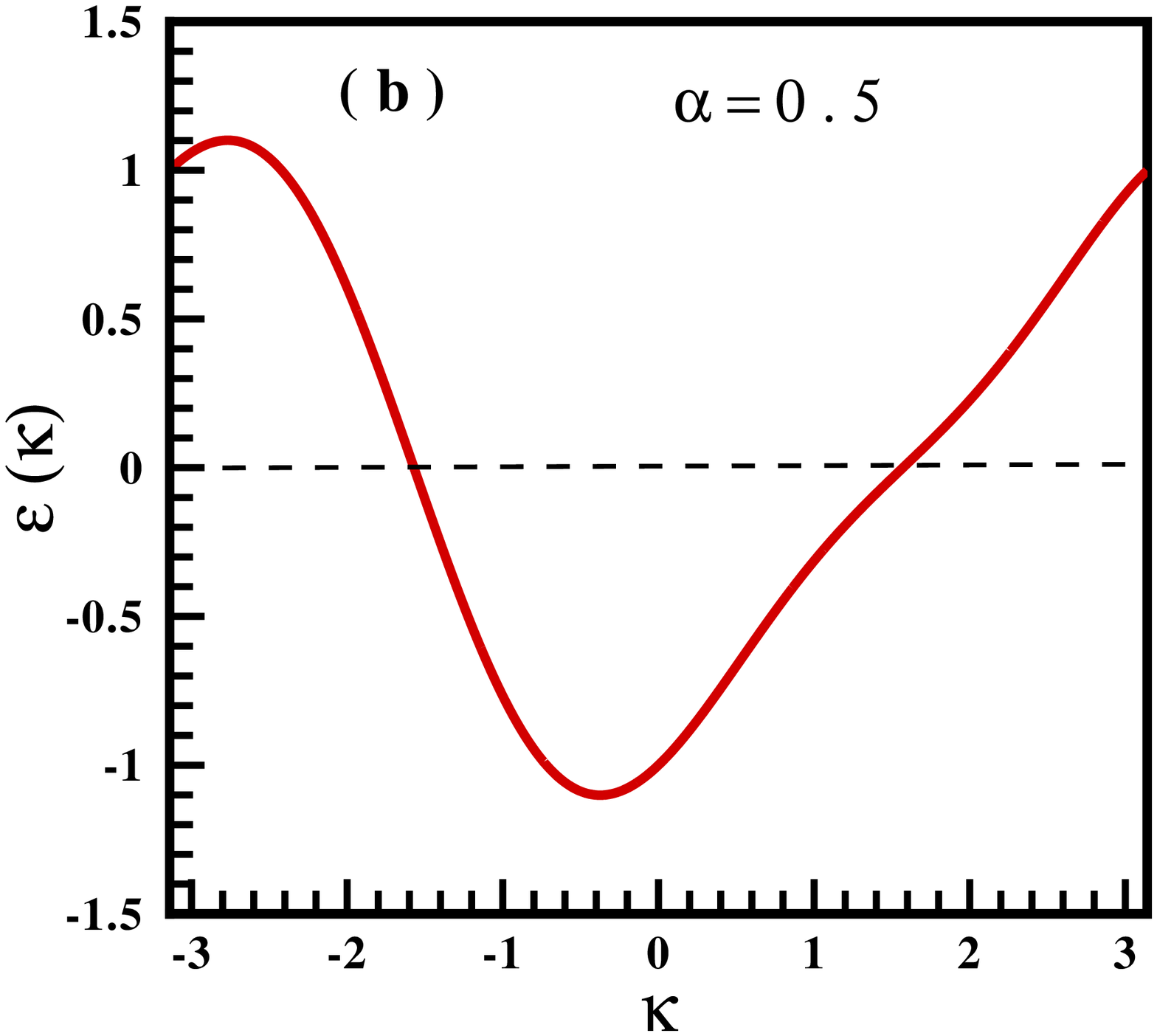,width=1.8 in}}
\centerline{\psfig{file=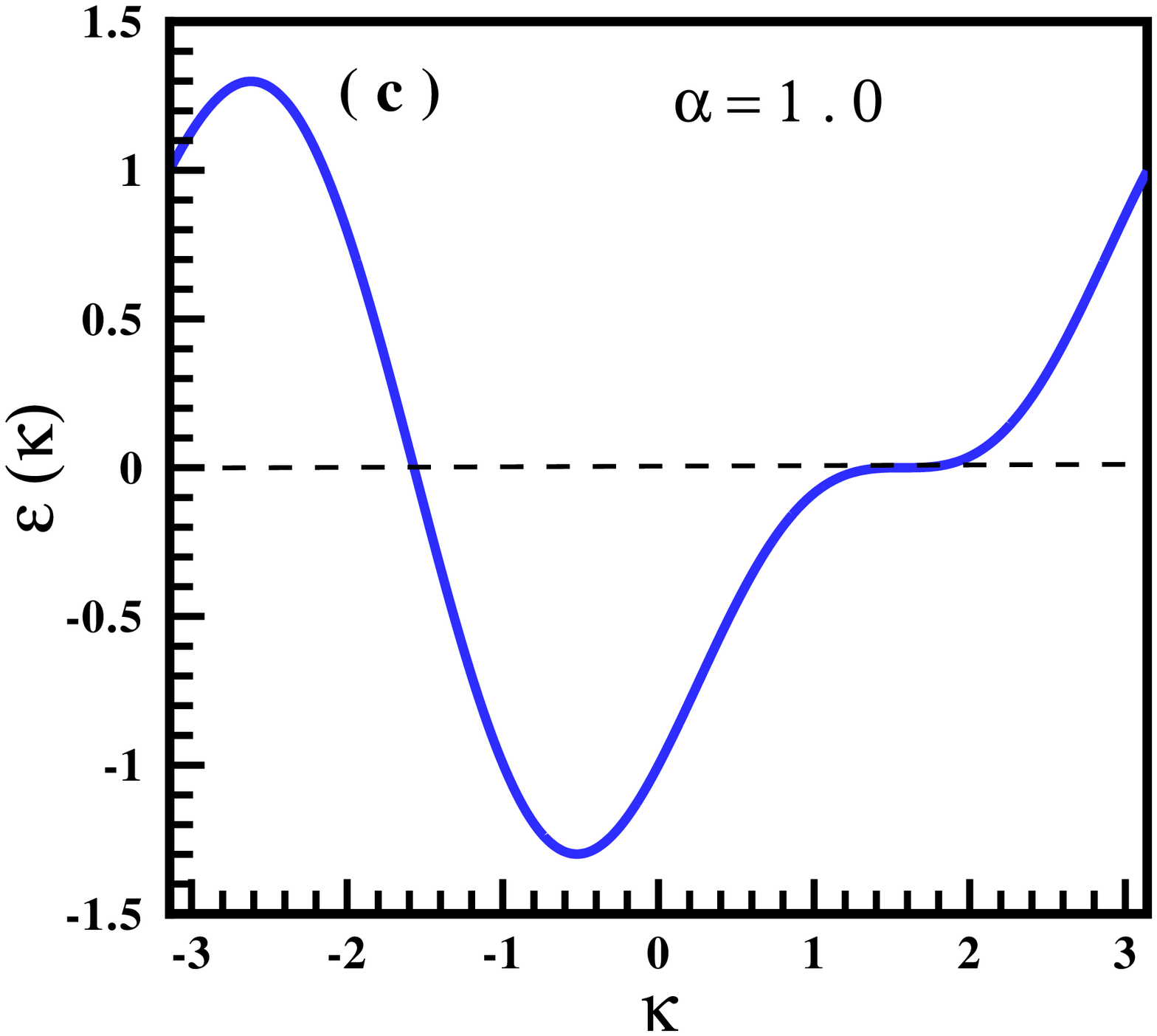,width=1.8 in}\psfig{file=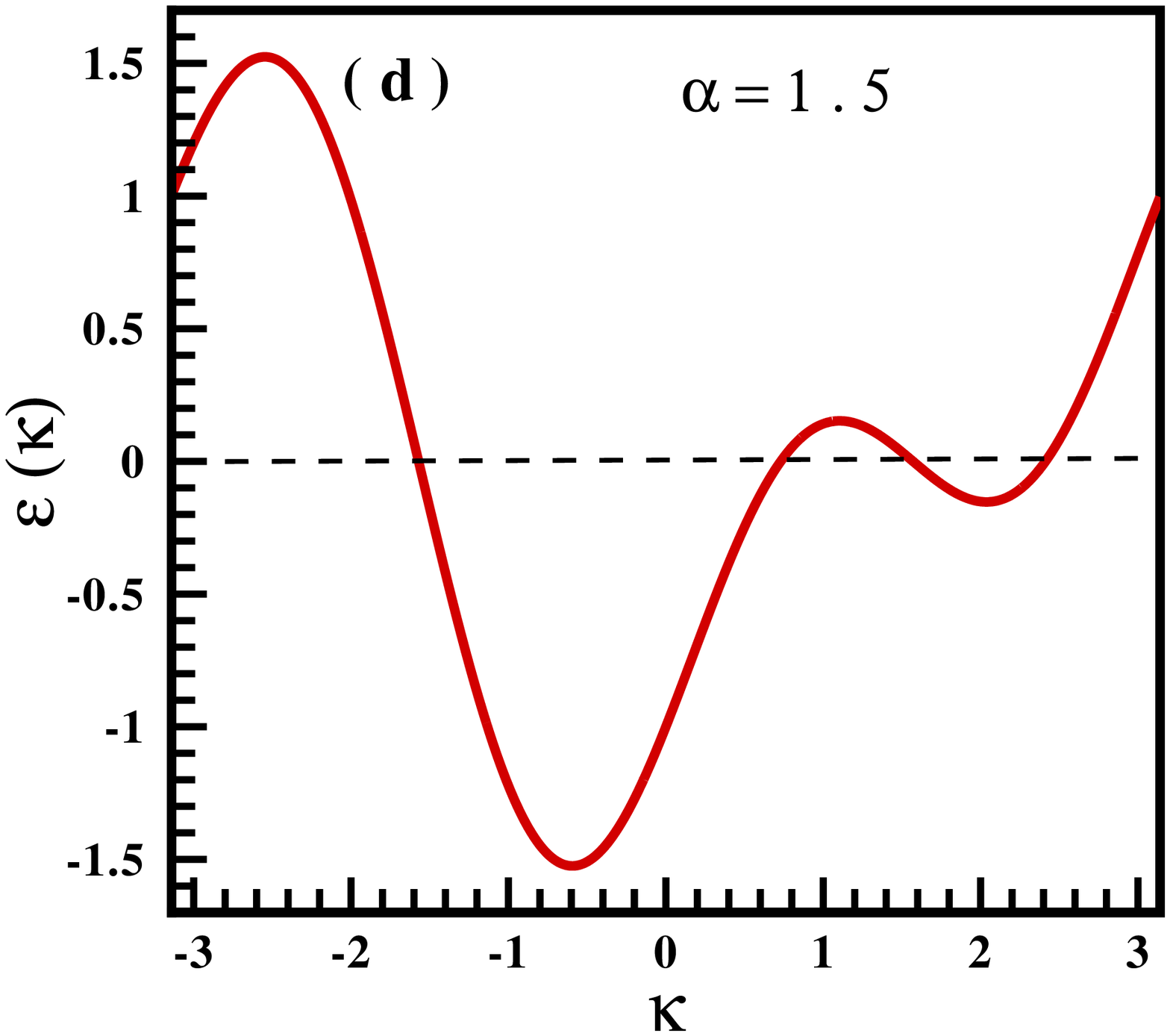,width=1.8 in}}
\caption{The Energy spectrum as a function of the momentum for different values of the TSI as:
 (a) In the absence of the TSI, $\alpha=0$, (b) In the presence of the TSI but less than quantum critical point, $\alpha=0.5<\alpha_c$, (c) In the presence of the TSI equal to the quantum critical point $\alpha=1.0=\alpha_c$, (d) In the presence of the TSI but larger than quantum critical point, $\alpha=1.5>\alpha_c$.}\label{fig66}
\end{figure}
It is known that the most important quantity in determining the dynamical behavior of the fermion systems is the energy spectrum $\varepsilon (k)$. One should note that the ground state of the chain model corresponds to the configuration where all states with $\varepsilon (k)<0$ are filled and $\varepsilon (k)>0$ are empty. In the region $\alpha<\alpha_c=1$, there is two-Fermi points at $k_F=\pm \frac{\pi}{2}$. Two additional Fermi points appear in the region $\alpha>\alpha_c=1$ at $k_F^{1}=\arcsin (\frac{1}{\alpha})$ and $k_F^{2}=\pi-\arcsin (\frac{1}{\alpha})$.  Since $\left\langle \psi_{GS} | \psi (t) \right\rangle=0$, the ground state energy has no effect in the dynamical behavior of the system (this is not general and depends on the initial state). Thus, only excited states of the Hamiltonian play important role in the dynamical behavior of the system. In the region $\alpha \leq \alpha_c=1 $, the ground state consists of all quasi-particles with momentum $-\frac{\pi}{2}<k<\frac{\pi}{2}$. But in the region  $\alpha > \alpha_c=1 $, it consists of  quasi-particles with momentum  $-\frac{\pi}{2}<k<k_{F}^{1}$ and $\frac{\pi}{2}<k<k_{F}^{2}$. Low-lying excited states of the chain model correspond to the configurations where in addition to filled states with $\varepsilon (k)<0$, quasi-particles with momentum value close to the Fermi points are also created. 

Here, we focus on the dispersion relation Eq.~(\ref{spec}) for different values of the TSI. Results are presented in Fig.~\ref{fig66}. In absence of the TSI, the asymmetry between quasi-particles with positive and negative momentum is seen in Fig.~(\ref{spec}) (a).  But, in presence of TSI (Fig.~\ref{fig66} (b-d)), the mentioned symmetry is broken and at first, the sub-branch with positive momentum ($k>k_F=\frac{\pi}{2}$) will be filled. As soon as the TSI crosses the quantum critical point, two additional Fermi points $k_{F}^{1(2)}$ are induced. In this case, quasi-particles with momentum $k_{F}^{1}<k<\frac{\pi}{2}$ will be also created in low-lying excited states and their impact as a signal shows itself in quantities such as the long-time average of the entanglement and the measure of the non-Markovianity. 
\\
\\

\section{Conclusion}\label{sec5}

In this paper, we have studied the dynamical behavior of a one-spin open quantum system in a one-dimensional spin-1/2 isotropic XX Heisenberg model with TSI. The TSI breaks $\pi/2$-rotation along $Z$-axis symmetry. We have shown that notable changes in the dynamics of the systems have occurred since the dynamics of the system is governed by Hamiltonian. Calculating the witness of non-Markovianity and the entanglement between the one-spin open quantum system and its environment, we showed that the one-spin open quantum system experiences no dynamical transition, since the one-way flow of information from the system to the environment was never observed. 

In the short-time dynamics of the entanglement between the one-spin and environment, we found a special time where the open quantum system becomes completely entangled with its environment. The entanglement-time shows decreasing behavior with respect to the TSI. We also calculated the long-time average of the entanglement and a clear signature of the quantum critical point observed. The mentioned signal of the quantum critical point observed also in the dynamical behavior of the measure of non-Markovianity.
\vspace{0.3cm}

\end{document}